\documentclass[aps,prl,twocolumn,superscriptaddress]{revtex4}

\usepackage{amssymb,amsmath,graphicx,graphics,color,array,caption,epstopdf}

\begin{document}

\title{Reduction of nonlinear field theory equations to envelope models: towards a universal understanding of analogues of relativistic systems}
\author{Charles W. Robson}
\affiliation{Laboratory of Photonics, Physics Unit, Tampere University, Tampere, FI-33720 Finland}
\author{Fabio Biancalana}
\affiliation{School of Engineering and Physical Sciences, Heriot-Watt University, EH14 4AS Edinburgh, UK}


\begin{abstract}
We investigate a novel mapping between solutions to several members of the Klein-Gordon family of equations and solutions to equations describing their reductions via the slowly varying envelope approximation. This mapping creates a link between the study of interacting relativistic fields and that of systems more amenable to laboratory-based analogue research, the latter described by nonlinear Schr{\" o}dinger equations. A new evolution equation is also derived, emerging naturally from the sine-Gordon formula, possessing a Bessel-function nonlinearity; numerical investigations show that solutions to this novel equation include quasi-solitary waves, breather solutions, along with pulse splittings and recombinations.
\end{abstract}


\maketitle

\paragraph{Introduction. ---} 



In the past few decades, the importance of the use of analogues in physics has been emphasised. These laboratory-based analogue experiments, designed to be described by the same equations as fundamental physical processes, are greatly useful due to the considerable experimental difficulties that can accompany the study of certain physical systems in their original contexts -- e.g. in the study of astrophysical black holes \cite{Unruh,Barcelo,Volovik,Volovik2,Steinhauer,Calum,Rousseaux,Visser_Silke,our_paper_1,our_paper_2}. A greater understanding of the relationships between solutions to certain nonlinear relativistic field theory equations (describing fundamental physics \cite{Landau1,Blundell1,Itzykson}) and solutions to those equations' reductions to more amenable formulae (which can describe analogues) would be especially advantageous.

In this work, we study the reductions, using the slowly varying envelope approximation (SVEA), of several nonlinear Klein-Gordon (KG) equations to nonlinear-Schr{\" o}dinger (NLS) equations, finding that a mapping exists between certain solutions to each. We believe that this may be of use in future analogue studies, as NLS equations are known to describe various systems which can be studied in the laboratory, including in optics \cite{Agrawal1,Ablowitz1,Ablowitz2}. Nonlinear KG equations, describing interacting relativistic fields, on the other hand describe much more extreme systems, less amenable to laboratory-based study. (For an example of a Bose-Einstein condensate analogue of the massive \emph{noninteracting} KG equation, see \cite{Visser_Silke}.)

Also presented in this work is a novel formula arising directly from the well-known sine-Gordon equation. We numerically model its evolutionary effects on several different fields, showing interesting phenomena, including wave splittings, recombinations and breathers (in some cases, with such low-amplitude breathing that they could be described as quasi-solitary waves).

\paragraph{Polynomial nonlinearities. ---} In this first section, we investigate three nonlinear Klein-Gordon equations \cite{Benci} that describe nonlinearities of a polynomial form, namely the cubic KG equation, the double-well KG equation, and finally the cubic-quintic KG equation.

The cubic nonlinear Klein-Gordon equation is given by
\begin{equation} \label{eq:CKG1}
\Box \phi + \lambda \phi^3 = 0.
\end{equation}
where $\Box$ is the d'Alembertian operator with signature (+ - - -). In all the results presented below, temporal and spatial derivatives are represented by $\dot{f}$ and $f'$, respectively.
Solutions for Eq. (\ref{eq:CKG1}) in 1+1 dimensions, in terms of Jacobi elliptic functions \cite{Abra}, include
\begin{equation} \label{eq:KG_sols1}
\begin{gathered}
\phi = \pm ic \sqrt{\frac{2}{\lambda}}\mathrm{sn}\left( cx \middle| -1 \right), \\
\phi = \pm \frac{ic}{\sqrt{\lambda}} \mathrm{cn}\left( cx \middle| 1/2 \right) , \\
\phi = \pm c \sqrt{\frac{2}{\lambda}}\mathrm{dc}\left( cx \middle| -1 \right) .
\end{gathered}
\end{equation}
To reduce the KG equations to the more tractable NLS equations (useful in optics, BEC theory, and other areas) we apply the slowly varying envelope approximation \cite{Agrawal1} to Eq. (\ref{eq:CKG1}). Rewriting the field in terms of its envelope, $\phi=(1/2) \left[ \psi \mathrm{exp}(-i\omega t) + \psi^{*} \mathrm{exp}(i\omega t) \right]$, and applying the SVEA leads directly to
\begin{equation} \label{eq:SW2}
i\dot{\psi}+\frac{1}{\omega}\psi'' - \frac{\lambda}{\omega} |\psi|^2 \psi = 0.
\end{equation}
(Parameters $\omega$ and the coupling $\lambda$ have been rescaled.) Three Jacobi elliptic function solutions in (1+1)d to Eq. (\ref{eq:SW2}) are given by
\begin{equation} \label{eq:SW3}
\begin{gathered}
\psi = \pm c \sqrt{\frac{2m}{\lambda}}\mathrm{sn}\left( cx \middle| m\right) e^{-i\frac{(1+m)c^2}{\omega}t} \quad (\mathrm{m>0}), \\
\psi = \pm ic \sqrt{\frac{2m}{\lambda}}\mathrm{cn}\left( cx \middle| m\right) e^{-i\frac{(1-2m)c^2}{\omega}t} \quad (\mathrm{m<0}), \\
\psi = \pm c\sqrt{\frac{2}{\lambda}}\mathrm{dc}\left( cx \middle| m\right) e^{-i\frac{(1+m)c^2}{\omega}t}.
\end{gathered}
\end{equation}

The core analytical result of our work, the mapping, emerges from the answer to the following question: how are solutions (\ref{eq:KG_sols1}) and (\ref{eq:SW3}) related? Firstly, for solutions (\ref{eq:SW3}) to become those of (\ref{eq:KG_sols1}), it is clear that the complex exponential factors in (\ref{eq:SW3}) must vanish. If we want both time $t$ and the parameter $c$ to be non-zero, then the value of $m$ must be constrained such that the complex exponential becomes one, i.e. its argument becomes zero. Explicitly, the first solution in (\ref{eq:SW3}) becomes the first solution in (\ref{eq:KG_sols1}) if parameter $m$ is set to $-1$, the second solution maps when $m=1/2$, and the third solution maps with $m=-1$. Note that the mappings of the sn and cn solutions in (\ref{eq:SW3}) occur when $m$ is fixed to integer values \emph{not in the solution domains}, e.g. the first solution in (\ref{eq:SW3}) is supported for $m>0$, but the mapping occurs at $m=-1$. The last solution in (\ref{eq:SW3}) can be mapped for $m=-1$, which is in the domain of solutions, so in this case both equations involved in the mapping are satisfied simultaneously.

\begin{figure}[h]
\centering
\includegraphics[scale=0.4]{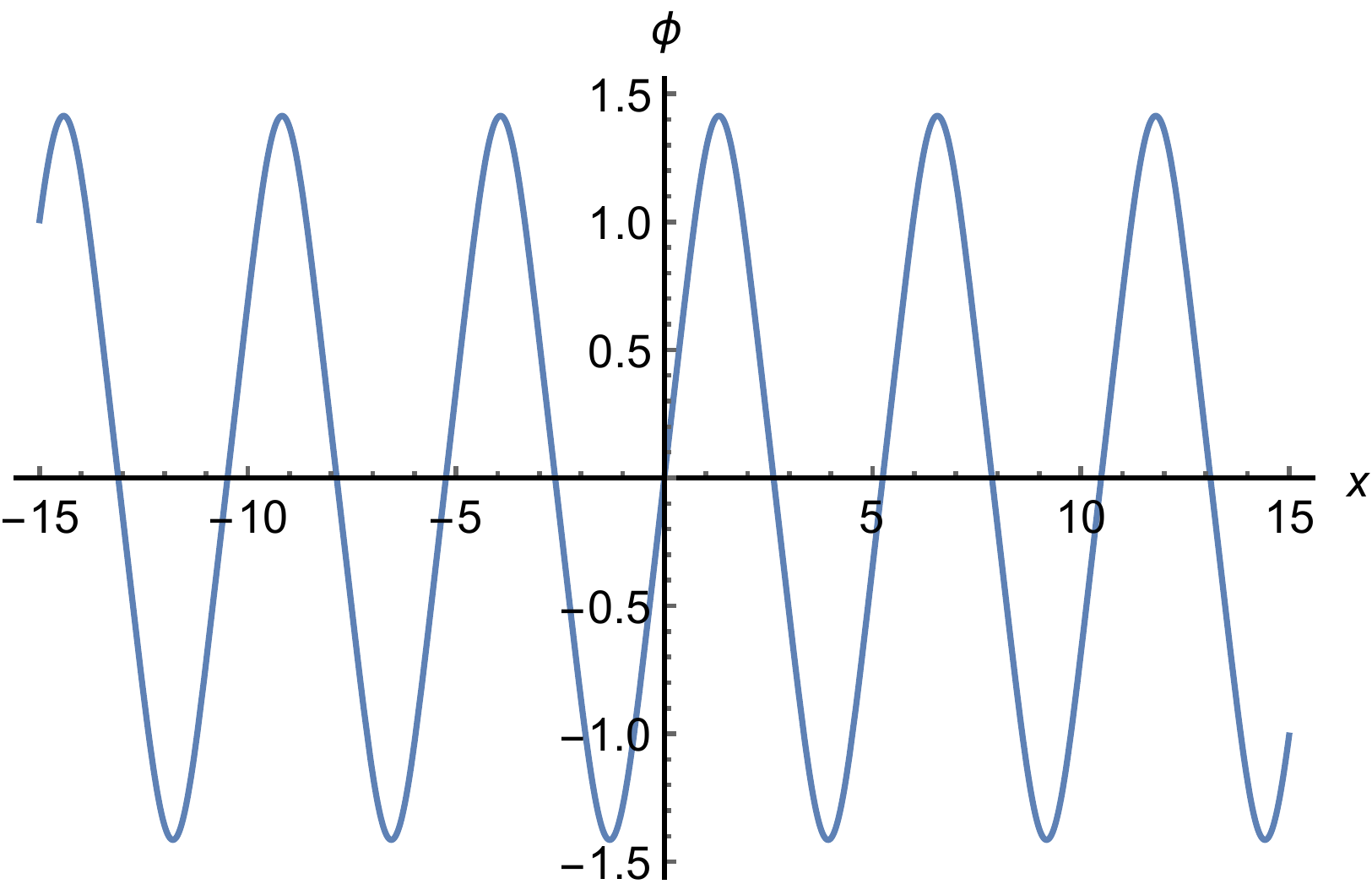}
\caption{Imaginary part of a periodic solution to the cubic KG equation: $\mathrm{Im}(\phi)=\sqrt{2}\mathrm{sn}\left( x \middle| -1 \right)$.}
\label{fig:cubic_SN}
\end{figure}

\begin{figure}[h]
\centering
\includegraphics[scale=0.4]{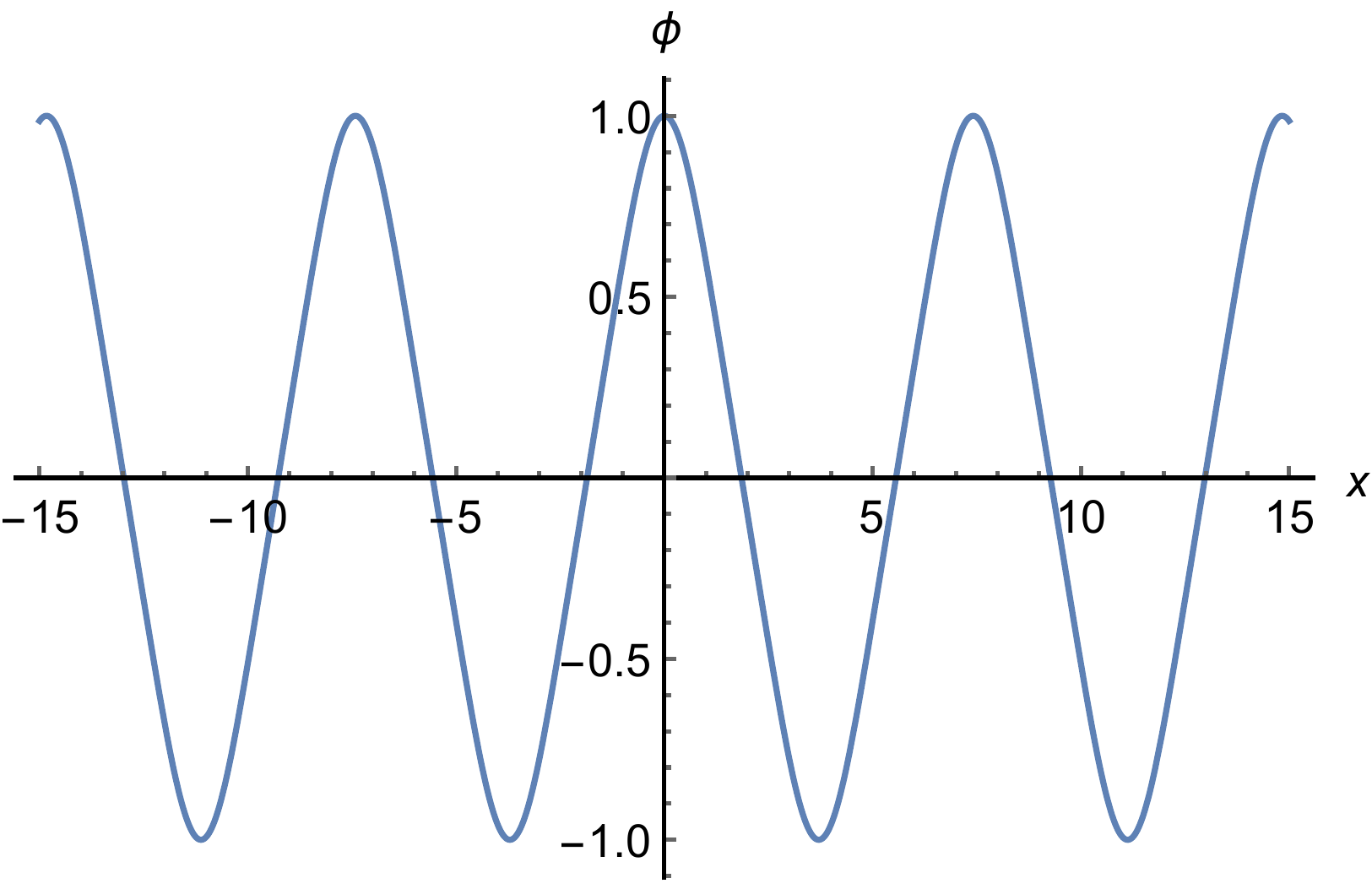}
\caption{Imaginary part of a periodic solution to the cubic KG equation: $\mathrm{Im}(\phi)=\mathrm{cn}\left( x \middle| 1/2 \right)$.}
\label{fig:cubic_CN}
\end{figure}

\begin{figure}[h]
\centering
\includegraphics[scale=0.4]{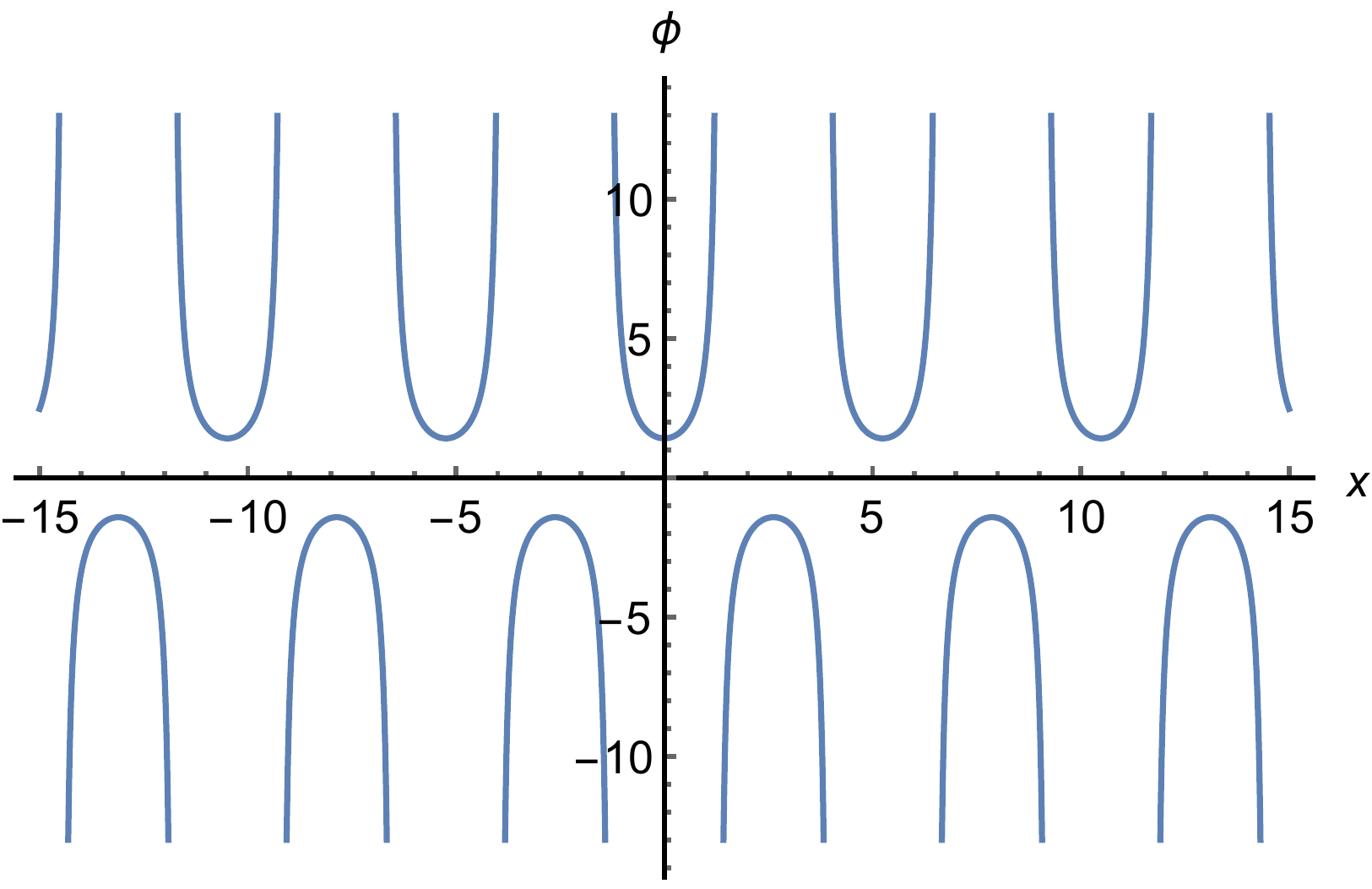}
\caption{A solution to the cubic KG equation: $\phi=\sqrt{2}\mathrm{dc}\left( x \middle| -1 \right)$.}
\label{fig:cubic_DC}
\end{figure}

Table~\ref{Table1} collects the solutions to both equations studied here, which are equal after parameter fixing. The $\phi$ solutions are plotted in Figs. \ref{fig:cubic_SN}, \ref{fig:cubic_CN} and \ref{fig:cubic_DC}, where $c$ and $\lambda$ have been set to $1$, and the $+$ solutions are chosen wherever a $\pm$ sign is present. The sn and cn solutions are purely imaginary and so the imaginary parts are plotted.

We look now at the mapping for a selection of solutions for the double-well system, which is governed by the equation
\begin{equation} \label{eq:DW0}
\Box \phi -m_s^2 \phi + \lambda \phi^3 = 0.
\end{equation}
It has kink/antikink solutions in (1+1)d:
\begin{equation} \label{eq:kink}
\phi=\pm\frac{m_s}{\sqrt{\lambda}}\mathrm{tanh}\left( \pm \frac{m_s}{\sqrt{2}}(x-x_0) \right);
\end{equation}
and other solutions, including
\begin{equation} \label{eq:DW1}
\begin{gathered}
\phi=\pm\sqrt{\frac{2}{\lambda}}\sqrt{m_s^2 - c^2}\mathrm{sn}\left( cx \middle| \frac{m_s^2 - c^2}{c^2} \right), \\
\phi=\pm i\sqrt{\frac{c^2 - m_s^2}{\lambda}}\mathrm{cn}\left( cx \middle| \frac{c^2 - m_s^2}{2c^2} \right), \\
\phi=\pm c \sqrt{\frac{2}{\lambda}}\mathrm{dc}\left( cx \middle| \frac{m_s^2 - c^2}{c^2} \right), \\
\phi= \pm \frac{m_s}{\sqrt{\lambda}} .
\end{gathered}
\end{equation}
Applying the SVEA to Eq. (\ref{eq:DW0}) yields
\begin{equation} \label{eq:DW2}
i\dot{\psi}+\frac{1}{\omega}\psi'' - \frac{\lambda}{\omega}\left( |\psi|^2 -\frac{m_s^2}{\lambda} \right) \psi = 0,
\end{equation}
where again we have rescaled parameters. This has solutions in (1+1)d:
\begin{equation} \label{eq:DW3}
\begin{gathered}
\psi = \pm c\sqrt{\frac{2}{\lambda}}\mathrm{tanh}(\pm c(x-x_0))e^{i\left( \frac{m_s^2 - 2c^2}{\omega} \right)t}, \\
\psi = \pm c \sqrt{\frac{2m}{\lambda}}\mathrm{sn}\left( cx\middle| m \right) e^{i \left( \frac{m_s^2 - (1+m)c^2}{\omega} \right) t} \quad (\mathrm{m>0}), \\
\psi = \pm i c \sqrt{\frac{2m}{\lambda}}\mathrm{cn}\left( cx \middle| m\right) e^{i \left( \frac{m_s^2 - (1-2m)c^2}{\omega} \right) t} \quad (\mathrm{m<0}), \\
\psi = \pm c\sqrt{\frac{2}{\lambda}}\mathrm{dc}\left( cx \middle| m\right) e^{i \left( \frac{m_s^2 - (1+m)c^2}{\omega} \right) t}, \\
\psi = \pm a e^{i \left( \frac{m_s^2 - a^2 \lambda}{\omega} \right) t}.
\end{gathered}
\end{equation}

\begin{figure}[h]
\centering
\includegraphics[scale=0.4]{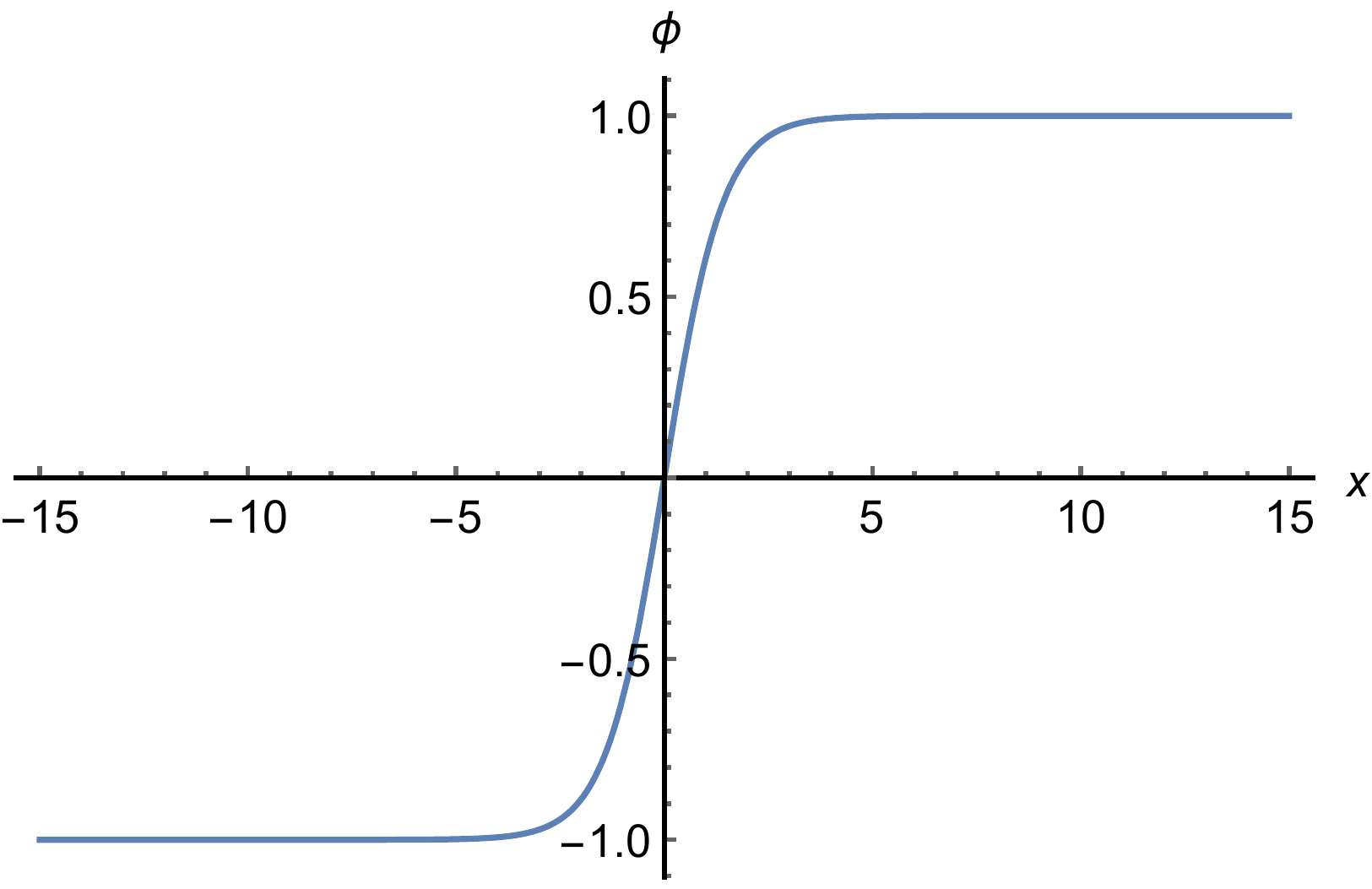}
\caption{A kink solution to the double-well KG equation: $\phi=\mathrm{tanh}\left( x / \sqrt{2} \right)$.}
\label{fig:DW1}
\end{figure}

Following the same pattern as seen above in the cubic case, all the solutions in (\ref{eq:DW3}) revert to those in (\ref{eq:kink}) and (\ref{eq:DW1}) when the parameter values are fixed to values that quench the time evolution. Table~\ref{Table2} collects the solution mappings. The first $\phi$ solution, a kink, is plotted in Fig. \ref{fig:DW1}, where $m_s$ and $\lambda$ have been set to $1$, $x_0$ has been set to $0$, and every $\pm$ sign is replaced with a $+$ sign.

Finally, we treat a more complicated nonlinearity, given by the cubic-quintic KG equation:
\begin{equation} \label{eq:CQ1}
\Box \phi - \sigma\phi^3 + \lambda\phi^5 = 0.
\end{equation}
This has a solution in (1+1)d of
\begin{equation} \label{eq:orig1}
\phi=\sqrt{\frac{3\sigma}{8\lambda}+\sqrt{\frac{3}{20\lambda}}\mathrm{sn}\left( x\middle| \frac{1}{5} \right)},
\end{equation}
where $15\sigma^2=16\lambda$. Applying the SVEA to (\ref{eq:CQ1}), we derive
\begin{equation}
i\partial_t \psi + \nabla^2 \psi + \sigma|\psi|^2 \psi - \lambda|\psi|^4 \psi=0,
\end{equation}
which has a solution in (1+1)d given by
\begin{equation} \label{eq:orig2}
\psi=\sqrt{\frac{3\sigma}{8\lambda}+\sqrt{\frac{3m}{4\lambda}}\mathrm{sn}\left( x\middle| m \right) }\mathrm{exp}\Big[ -i\left( \frac{1+m-\frac{9\sigma^2}{8\lambda}}{4} \right) t \Big],
\end{equation}
where $16m\lambda=3\sigma^2$. As we saw before for the single- and double-well equations, this solution maps, i.e. (\ref{eq:orig2}) maps to (\ref{eq:orig1}), once the former's time variation component vanishes. The component will vanish iff $1+m-(9\sigma^2 / 8\lambda )=0$, assuming non-zero $t$. This occurs when $m=1/5$, again leading to a mapping between separate solutions.


We posit that our mapping procedure is possible only when solutions are explicitly separable into spatial and temporal parts (such as in the $\psi$ solutions shown above with an amplitude depending only on $x$ and a periodic component depending only on $t$). An example where this lack of separability prevents a mapping is the (2+1)d instantonic solution $\phi=\left( 3 / \lambda \right)^{1/4}\sqrt{\rho / (x^2 + y^2 + z^2 + \rho^2)}$ to the quintic KG equation, $\Box \phi + \lambda\phi^5 = 0$ (after a Wick rotation). This solution cannot be mapped simply, which we postulate is due to its nonseparable form. It is also important to note that the SVEA acting on a general formula could transform an integrable equation into a nonintegrable one, or vice versa, and hence applying the SVEA could ``create" or ``destroy" solutions, hindering certain mappings.

\begin{widetext}
\centering
\captionof{table}{Cubic Klein-Gordon Mappings} 
\label{Table1}
{\renewcommand{\arraystretch}{2.6}
\begin{tabular}{ |p{7.5cm}|p{8.5cm}|  }
 \hline
 \qquad \qquad \qquad \ $\Box \phi + \lambda \phi^3 = 0$ solutions & \qquad \quad \ $i\dot{\psi}+\frac{1}{\omega}\psi'' - \frac{\lambda}{\omega} |\psi|^2 \psi = 0$ solutions \\
 \hline \hline
 \qquad \qquad \ $\phi = \pm ic \sqrt{\frac{2}{\lambda}}\mathrm{sn}\left( cx \middle| -1\right)$   & \qquad \ $\psi = \pm c \sqrt{\frac{2m}{\lambda}}\mathrm{sn}\left( cx \middle| m\right) e^{-i\frac{(1+m)c^2}{\omega}t}, \quad (\mathrm{m>0})$ \\
 \hline
 \qquad \qquad \ $\phi = \pm \frac{ic}{\sqrt{\lambda}} \mathrm{cn}\left( cx \middle| \frac{1}{2}\right)$ & \qquad \ $\psi = \pm ic \sqrt{\frac{2m}{\lambda}}\mathrm{cn}\left( cx \middle| m\right) e^{-i\frac{(1-2m)c^2}{\omega}t}, \quad (\mathrm{m<0})$ \\
 \hline
 \qquad \qquad \ $\phi = \pm c \sqrt{\frac{2}{\lambda}}\mathrm{dc}\left( cx \middle| -1\right)$ &   \qquad \ $\psi = \pm c\sqrt{\frac{2}{\lambda}}\mathrm{dc}\left( cx \middle| m\right) e^{-i\frac{(1+m)c^2}{\omega}t}$ \\
 \hline
\end{tabular}}

\clearpage

\centering
\captionof{table}{Double-Well Klein-Gordon Mappings}
\label{Table2}
{\renewcommand{\arraystretch}{2.6}
\begin{tabular}{ |p{7.5cm}|p{8.5cm}|  }
 \hline
 \qquad \qquad \ \ $\Box \phi -m_s^2 \phi + \lambda \phi^3 = 0$ solutions & \qquad $i\dot{\psi}+\frac{1}{\omega}\psi'' - \frac{\lambda}{\omega}\left( |\psi|^2 -\frac{m_s^2}{\lambda} \right) \psi = 0$ solutions \\
 \hline \hline
 \qquad $\phi=\pm\frac{m_s}{\sqrt{\lambda}}\mathrm{tanh}\left( \pm \frac{m_s}{\sqrt{2}}(x-x_0) \right)$   & \qquad $\psi = \pm c\sqrt{\frac{2}{\lambda}}\mathrm{tanh}(\pm c(x-x_0))e^{i\left( \frac{m_s^2 - 2c^2}{\omega} \right)t}$ \\
 \hline
 \qquad $\phi=\pm\sqrt{\frac{2}{\lambda}}\sqrt{m_s^2 - c^2}\mathrm{sn}\left( cx \middle| \frac{m_s^2 - c^2}{c^2} \right)$ &   \qquad $\psi = \pm c \sqrt{\frac{2m}{\lambda}}\mathrm{sn}\left( cx \middle| m\right) e^{i \left( \frac{m_s^2 - (1+m)c^2}{\omega} \right) t}, \quad (\mathrm{m>0})$ \\
 \hline
 \qquad $\phi=\pm i\sqrt{\frac{c^2 - m_s^2}{\lambda}}\mathrm{cn}\left( cx \middle| \frac{c^2 - m_s^2}{2c^2} \right)$ &   \qquad $\psi = \pm i c \sqrt{\frac{2m}{\lambda}}\mathrm{cn}\left( cx \middle| m\right) e^{i \left( \frac{m_s^2 - (1-2m)c^2}{\omega} \right) t}, \quad (\mathrm{m<0})$ \\
 \hline
 \qquad $\phi=\pm c \sqrt{\frac{2}{\lambda}}\mathrm{dc}\left( cx \middle| \frac{m_s^2 - c^2}{c^2} \right)$ &   \qquad $\psi = \pm c\sqrt{\frac{2}{\lambda}}\mathrm{dc}\left( cx \middle| m\right) e^{i \left( \frac{m_s^2 - (1+m)c^2}{\omega} \right) t}$ \\
 \hline
 \qquad $\phi= \pm \frac{m_s}{\sqrt{\lambda}}$ & \qquad $\psi = \pm a e^{i \left( \frac{m_s^2 - a^2 \lambda}{\omega} \right) t}$ \\
 \hline
\end{tabular}}
\newline \newline
\end{widetext}

\paragraph{Non-polynomial nonlinearities. ---} In this section, we look at a system with a particular non-polynomial nonlinearity, described by the ubiquitous ``sine-Gordon" equation \cite{Rajaraman}:
\begin{equation} \label{eq:SG1}
\Box \phi + \lambda^2\rm{sin}\phi = 0.
\end{equation}
This has a kink solution in (1+1)d, given by
\begin{equation} \label{eq:kink2}
\phi = 4\mathrm{arctan}\left( e^{\lambda\gamma (x-\nu t) + \delta}\right),
\end{equation}
and a purely imaginary solution \cite{Wang1}:
\begin{equation} \label{eq:Wang}
\phi = \pm2\mathrm{arctan}\left( \pm i\mathrm{sn}\left( \pm \frac{x}{\sqrt{2}} \middle| -1 \right) \right),
\end{equation}
where $\lambda$ has been set to unity.

Expanding $\phi$ in terms of its envelope, as was our procedure in the previous section, transforms equation (\ref{eq:SG1}) into
\begin{equation}
i\dot{\psi}+\frac{1}{2\omega}\psi '' - \left( \frac{\lambda^2}{2\omega}\right) {}_0F_1\left( 2, -\frac{|\psi|^2}{4} \right)\psi = 0,
\end{equation}
where ${}_0F_1()$ is a hypergeometric function; alternatively this can be written in terms of a Bessel function of the first kind $J_1()$ as
\begin{equation} \label{eq:Bes1}
i\dot{\psi}+\frac{1}{2\omega}\psi '' - \lambda^2\frac{\psi}{\omega|\psi|} J_1 (|\psi|) = 0.
\end{equation}

An analytical solution to equation (\ref{eq:Bes1}) has yet to be found, despite the form of solutions (\ref{eq:kink2}) and (\ref{eq:Wang}) which suggests a mapping may be possible; we speculate that the lack of a solution could be due to the mapping being not bijective. Due to the lack of analytical solutions to Eq. (\ref{eq:Bes1}), we have performed numerical simulations to study the evolution of several different initial waves under the influence of this equation. (As an aside, when $\psi$ has no spatial dependence and is of the form $\psi=\psi_0 \mathrm{exp}(ikt)$, then formula (\ref{eq:Bes1}) does have the following basic solution: $\psi=\psi_0 \mathrm{exp}\left( -i \lambda^2 J_1(\psi_0) t / \omega \psi_{0} \right)$.

\begin{figure}[h]
\centering
\includegraphics[scale=0.4]{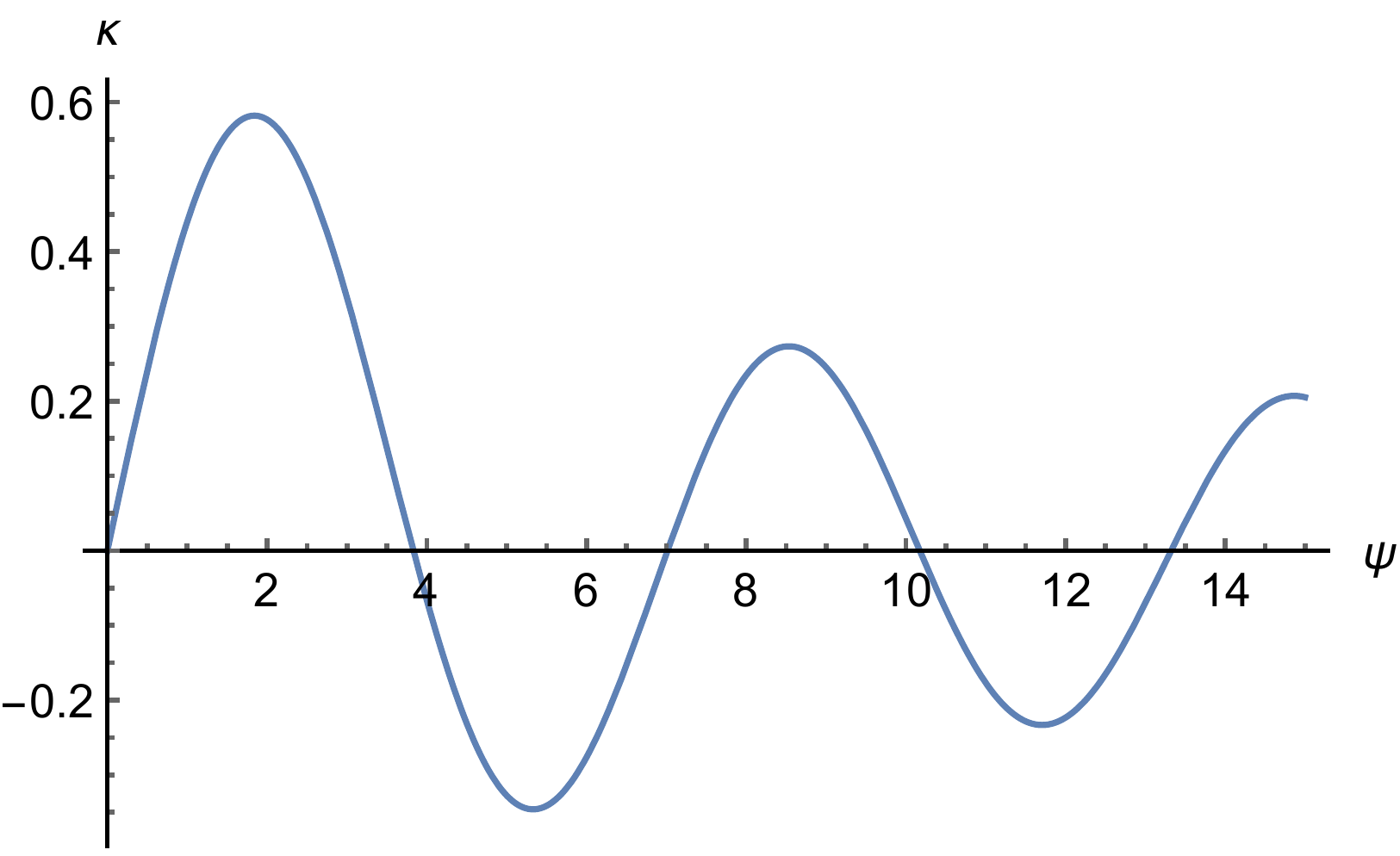}
\caption{Oscillation of equation (\ref{eq:Bes1})'s nonlinear parameter $\kappa$ (after redefinition).}
\label{fig:Bessel}
\end{figure}

Let us look more closely at the nonlinear term in equation (\ref{eq:Bes1}), as it is very different from those in our previous examples. By comparing our formula with the nonlinear Schr{\" o}dinger equation $i\dot{\psi}+(1/2)\psi '' - \kappa |\psi|^2 \psi = 0$, we find that the parameter $\kappa$ in our case is $\kappa=\left[ J_1 (|\psi|)/|\psi|^3 \right]$ (setting $\omega=\lambda=1$). When $\kappa<0$ we are in the focusing regime -- in which bright solitons can form. Fig. \ref{fig:Bessel} shows a plot of $\kappa=J_1 (|\psi|)$ vs. $\psi$ (in the plot, the $|\psi|^{-3}$ factor has been ignored and so $\kappa$ redefined, because the term doesn't affect the pertinent features, i.e. the sign of the nonlinearity or the location of its zeros). The first zero of $\kappa$ occurs at the first Bessel zero, $\psi \approx 3.8317$, so a wave below this threshold amplitude will be completely in the defocusing regime (although, each part of the wave would have a different amplitude and so would experience differing magnitudes of $\kappa$). As can be seen, a pulse of amplitude greater than $\sim$14, in the same units as in Fig. \ref{fig:Bessel}, would experience at least five different nonlinear regimes, resulting in a highly non-trivial evolution.

\begin{figure}[h]
\centering
\includegraphics[scale=0.4]{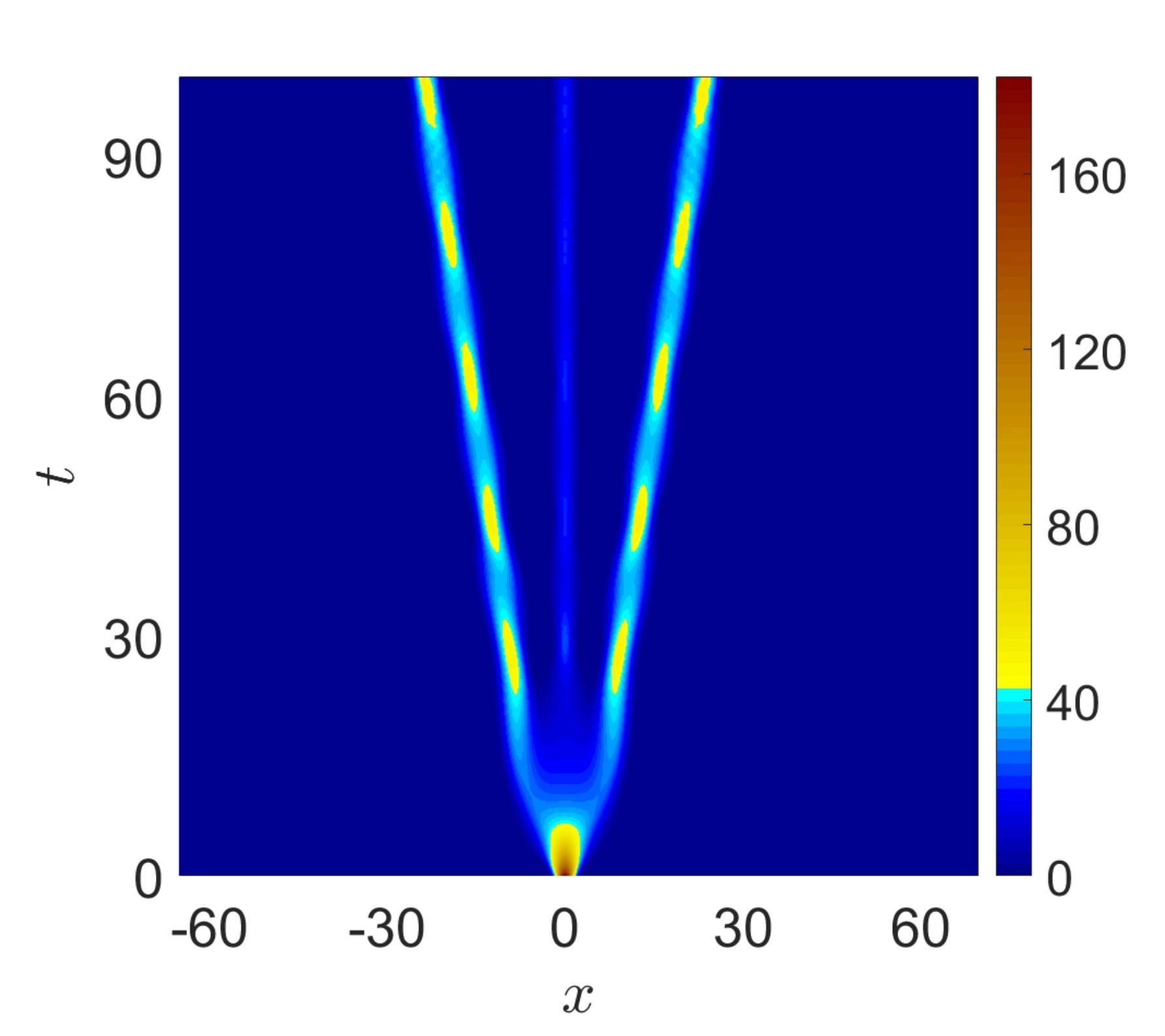}
\caption{Evolution of initial pulse $\psi=15\mathrm{sech}\left( x \right)$.}
\label{fig:Wave1}
\end{figure}

\begin{figure}[h]
\centering
\includegraphics[scale=0.4]{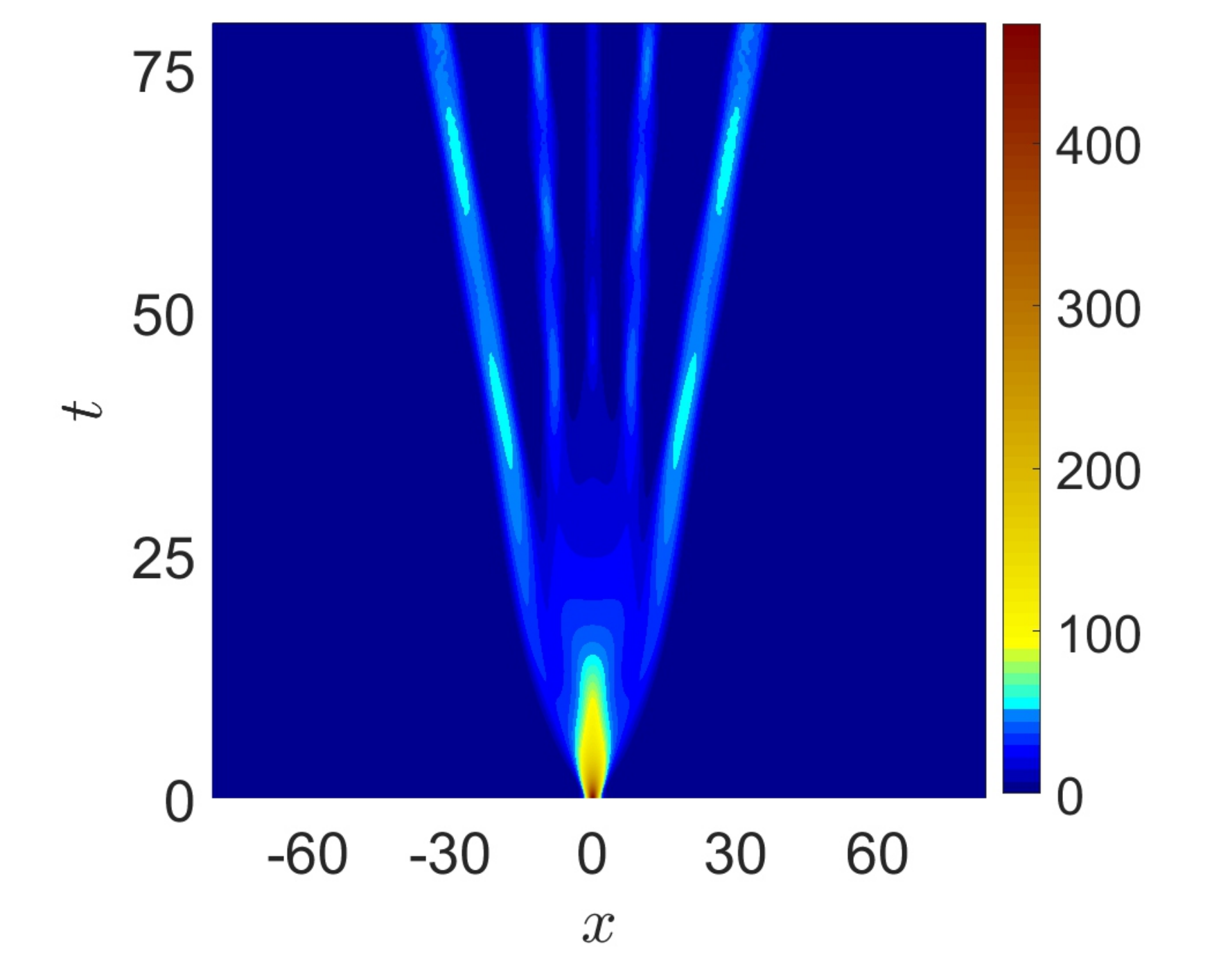}
\caption{Evolution of initial pulse $\psi=22\mathrm{sech}\left( x \right)$.}
\label{fig:Wave2}
\end{figure}

\begin{figure}[h]
\centering
\includegraphics[scale=0.4]{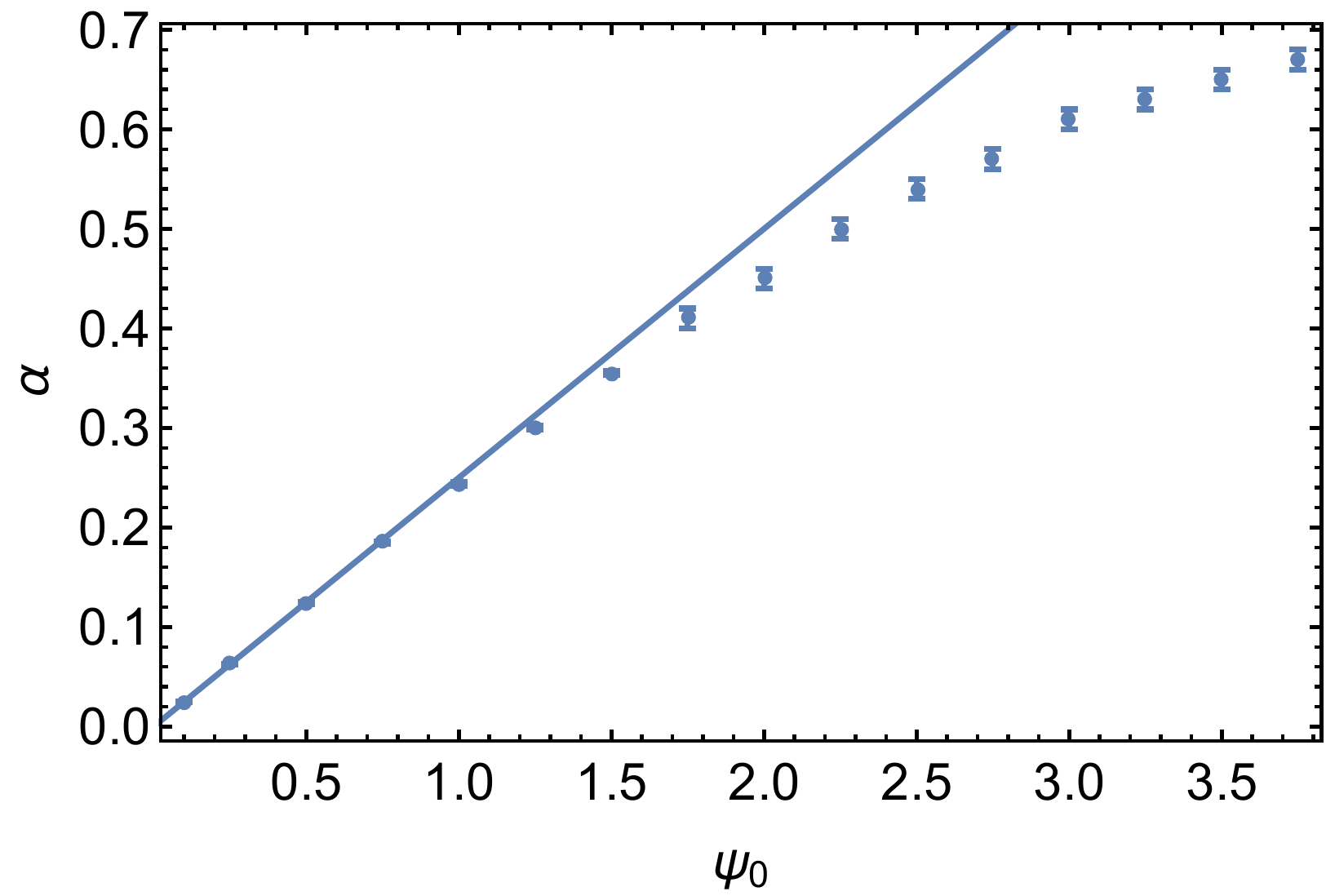}
\caption{Data points showing the values of $\psi_{0}$ and $\alpha$ at which $\psi=\psi_{0}\mathrm{sech}\left(\alpha x \right)$ is (almost) unchanged throughout its evolution (it evolves as a small-amplitude standing breather). The data points are numerical results, whereas the solid line $\psi_{0}=4\alpha$ is derived from theory and is valid only for small $\psi$.}
\label{fig:stable}
\end{figure}

\begin{figure}[h]
\centering
\includegraphics[scale=0.4]{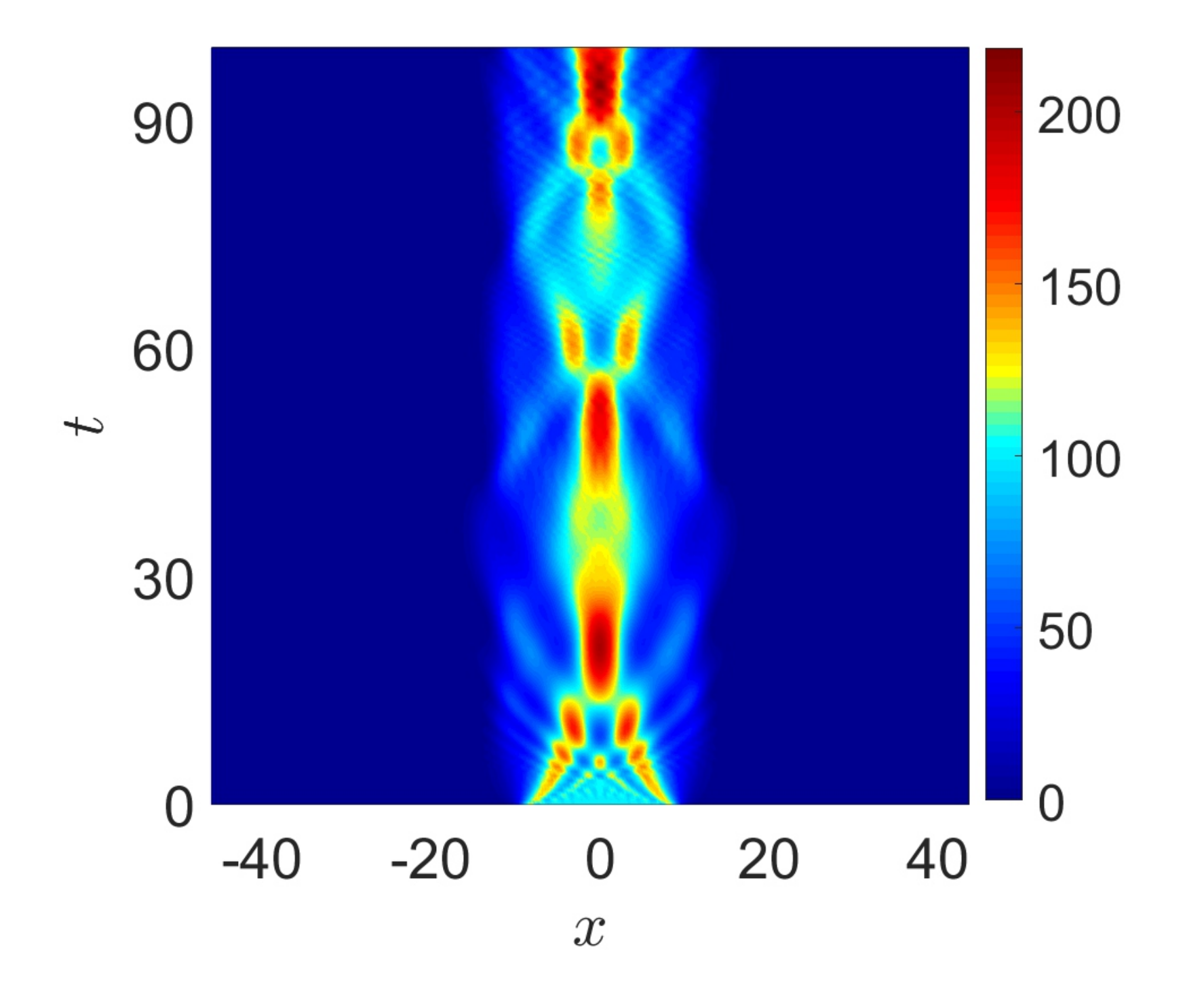}
\caption{Evolution of initial pulse $\psi=10\mathrm{exp}\left( -(x/10)^{40} \right)$.}
\label{fig:Wave3}
\end{figure}

Below, we present numerical results of several different initial waves evolving through formula (\ref{eq:Bes1}). The sech-shaped pulses will be parameterised $\psi=\psi_{0}\mathrm{sech}\left( \alpha x \right)$ with amplitude $\psi_{0}$ and width parameter $\alpha$. In each contour plot below, the numbers on the right-hand colour scale indicate $|\psi|^2$ values.

\emph{Case 1}: $\psi=15\mathrm{sech}\left( x \right)$.

The evolution of this wave contains several interesting features. As can be seen in the contour plot of Fig. \ref{fig:Wave1}, the pulse initially disperses, after which it splits into three localised structures: a central one maintaining its position whilst oscillating slightly (a standing breather), and two larger-amplitude structures moving apart and oscillating substantially (traveling breathers).

\emph{Case 2}: $\psi=22\mathrm{sech}\left( x \right)$.

If we now simulate a wave with a higher initial amplitude, a greater number of breathers are produced, as shown in Fig. \ref{fig:Wave2} -- one is standing and four are travelling apart in pairs. As in the previous case, the breathers exhibit greater amplitudes when they are present further out from the centre $x=0$, with the central standing breather having very small amplitude and oscillating only weakly.

\emph{Case 3}: $\psi=0.4\mathrm{sech}\left(0.1 x \right)$.

In this case, the wave keeps its initial shape indefinitely, except for having a very small oscillation -- it is a weakly-oscillating standing breather. This wave is a single case of a special class of solutions showing very stable behaviour. These solutions live on a line of stability in ($\alpha$, $\psi_{0}$) space, a pattern that breaks down somewhat after the amplitude of the wave has passed the first Bessel zero (after this point, the oscillations become larger in amplitude and cannot be categorised as ``weak"). Taking a Taylor series of the Bessel nonlinearity of equation (\ref{eq:Bes1}), truncating at third order, and solving the equation for $\psi=\psi_{0}\mathrm{sech}\left( \alpha x \right)\mathrm{exp}(ikt)$, shows that a line of stable solutions exists for $\psi_{0}=4\alpha$, agreeing with our numerical results for $\psi=0.4\mathrm{sech}\left(0.1 x \right)$. This derivation is only valid for small $\psi$ and, as expected, deviates from the numerical results at higher amplitudes, as shown in Fig. \ref{fig:stable}.

\emph{Case 4}: $\psi=\psi_{0}\mathrm{exp}\left[ -(x/10)^{40} \right]$.

This wave is a step-like wave: a supergaussian of very high order. Fig. \ref{fig:Wave3} illustrates the evolution of $\psi=10\mathrm{exp}\left[ -(x/10)^{40} \right]$, showing features including periodic splittings and recombinations, as well as large sharp central peaks which periodically emerge and then recede. As is clear, the evolution of this initial wave is much more complicated and shows less explicit structure during propagation than the previous cases.

A theoretical analysis of the effects of this new equation -- Eq. (\ref{eq:Bes1}) -- on classes of different initial waves will be the subject of a future publication.

\paragraph{Conclusions. ---} We have found that mappings exist between several solutions to equations describing interacting relativistic fields, important in many areas of physics, and solutions to models found using the slowly varying envelope approximation. The number of mappable solutions could be vast, as there is no apparent impediment to our method being extended to many other systems, and we believe that our work is a useful first step in the construction of new analogues of field-theoretical phenomena. An intriguing new equation emerging directly from the sine-Gordon formula is also introduced in this work, with Bessel-function nonlinearity. It is shown numerically that it has a highly nontrivial effect on various waves -- including the creation of breathers and quasi-stable states. We hope that our findings will benefit future investigations in analogue physics and stimulate further analysis of the properties of the novel Schr{\" o}dinger-type equation with Bessel-function nonlinearity introduced here.

\end{document}